# Exchange bias and its phase reversal in the zero magnetization admixed rare earth intermetallics

Prasanna D. Kulkarni*, S. Venkatesh, A. Thamizhavel, V. C. Rakhecha, S. Ramakrishnan and A. K. Grover

Department of Condensed Matter Physics and Materials Science,

Tata Institute of Fundamental Research, Colaba, Mumbai-400005.

*email : prasanna@tifr.res.in

## Abstract

Exchange bias effect is an important attribute in several device applications. Traditionally, it is discussed as a form of exchange anisotropy at the interface between the ferromagnetic/antiferromagnetic layers of two distinct systems. We report here on the magnetically ordered alloys possessing the feature of unidirectional anisotropy, which reverses its sign across a characteristic temperature. These are admixed rare earth intermetallics, comprising two dissimilar rare earth (RE) ions, belonging to the two different halves of the 4$f$-series and imbibing the notion of ferromagnetic coupling between the spins of all 4$f$-rare earth ions and that of the conduction electrons. Magnetic moments of dissimilar RE ions in such admixed alloys, however, get antiferromagnetically coupled and they display magnetic compensation feature such that the field induced reversal in the orientation of the magnetic moments of dissimilar rare earth ions happens across the compensation temperature ($T_{\text{comp}}$). Above a threshold field, the conduction electron polarization also reverses its sign across $T_{\text{comp}}$, this is argued to correlate with the observed phase reversal in the exchange bias field.

**Introduction**

Exchange bias field is notionally identified as a shift in the magnetization hysteresis loop (*M-H*) from the origin along the field axis and it is typically observed in ferromagnetic-antiferromagnetic multilayer composites [1,2], as a given sample is cooled down in an external field below the Neel temperature ($T_N$) of the antiferromagnet. The materials with such an attribute have application potential in the areas of information storage technology, magnetic field sensors and MRAM devices [3-5]. We draw attention here to the admixed rare earth alloys [6-11] derived from a ferromagnetic series of isoelectronic rare earth intermetallics, and assert that these are an attractive class of materials for devices exploiting the notion of exchange bias field and its phase reversal across a specific temperature. Such alloys can be easily fabricated in different material forms and dimensions, and can hence be integrated into complex multi-layer composites for characteristic applications.

The fourteen rare earth (RE) elements (atomic number 58 to 71), Cerium to Lutetium, are included together with elemental Lanthanum (atomic number 57) in the periodic table for their similar chemical attributes due to isoelectronic ($6s^2$-$5d^1$) valence state and progressive filling of deeper lying 4*f* shell as per the Hund's rule and with strong spin-orbit coupling ($\lambda$ **L.S**) operating between resultant spin (**S**) and orbital (**L**) angular momentum, such that the total angular momentum, **J = L+S**. In metallic systems, 4*f*-spins are exchange coupled to the spins of conduction electrons via indirect RKKY interactions, which can result in either ferromagnetic or antiferromagnetic coupling [6,7,12] between the neighboring spins of rare earth ions in a given series of isoelectronic rare earth intermetallic compounds. A ferromagnetic rare earth compound can be driven to the zero-magnetization response limit by substitution of a suitable fraction of RE element by another RE element belonging to the opposite half of the 4*f*-series [6-11]. This is based on the premise that RKKY coupling maintains the ferromagnetic alignment between spins of all RE ions [6,13,14], whether they belong to first half or second half of the 4*f* series, where the eigen value of the total angular momentum (*J*) in the ground state equals *L* - *S* or *L* + *S*, respectively. The ferromagnetic coupling between the 4*f*-spins of the RE ions makes the magnetic moments ($<\mu_z> = -g_J\mu_B <J_z>$, where $g_J$ is the Lande's *g*-factor) of dissimilar RE ions in an admixed $R_{1-x}R'_x\text{Al}_2$ system (where *R* and *R'* belong to different halves of 4*f* series) antiferromagnetically aligned. In stoichiometries close to the notional zero magnetization limit (for instance, where, $x\mu_R \sim (1-x)\mu_{R'}$), one can witness quasi-ferrimagnetic behaviour. In several such ferrimagnets, at low applied fields one comes across magnetic compensation phenomenon between the responses originating from local moments of *R* and *R'*, and abetted by a small but important contribution from conduction electron polarization. It is well documented in the literature [8,9,11] that the imposition of a larger applied field can reverse the orientation of the aligned moments of $\mu_R$ and $\mu_{R'}$ across the magnetic compensation temperature, which in turn implies reversal in the sign of the contribution of CEP to the net magnetization (*M*) as well.

In recent years, there has been an interest in establishing the field-induced reversal in the orientation of RE moments as a phase transition [15]. After investigations in several ferromagnetic series of the admixed RE alloys, we now reckon that the phenomenon of reversal of magnetic moments in them imprints as an asymmetric shift in the magnetization hysteresis loops, which can be termed as an exchange bias field ($H_{\text{exch}}$). We find that the exchange bias field surfaces up as we get near the regions of the magnetic compensation temperature ($T_{comp}$), $H_{exch}$ is observed to rapidly rise as $T$ approaches $T_{comp}$, and across $T_{comp}$, it reverses its phase and diminishes rapidly again. The half-width of the magnetization hysteresis loop across the $M = 0$ axis, which is notionally referred to as an effective coercive field $H_c^{eff}$ [16] also registers a very characteristic behaviour across $T_{comp}$, viz., a divergence in $H_c^{eff}$ from low and high temperature sides well away from $T_{comp}$, followed by a collapse in the width of the $M$-$H$ (i.e., $H_c^{eff}$), as $T$ approaches $T_{comp}$ from the either end. We choose to present here the generic behaviour in three specific admixed alloys in the ferromagnetic RE-dialuminide series, viz., $Sm_{1-x}Gd_xAl_2$ ($x = 0.02$, polycrystalline sample), $Pr_{1-x}Gd_xAl_2$ ($x = 0.2$, polycrystalline sample) and $Nd_{1-x}Ho_xAl_2$ ($x = 0.25$, a single crystal sample). Various aspects of magnetic compensation phenomenon specific to $Sm^{3+}$ ions in Gd doped $SmAl_2$ alloys have received large attention during last ten years [9,11,15,17], whereas, the magnetic compensation in the other two systems had been identified in 1960s [6,7]. We shall highlight the generic connection between the magnetic compensation phenomenon and the exchange bias field in the series of admixed rare earth intermetallics.

**Experimental**

The polycrystalline samples of the admixed alloys, $Sm_{0.98}Gd_{0.02}Al_2$ and $Pr_{0.8}Gd_{0.2}Al_2$ were prepared by melting together in an arc furnace the appropriate amounts of the end members, $SmAl_2$/$PrAl_2$ and $GdAl_2$. The initially synthesized end members were first characterized by powder X-ray diffraction and confirmed to have cubic *C15* structure. The powdered samples of the admixed stoichiometries were also confirmed to be in cubic *C15* structure. The single crystal sample of $Nd_{0.75}Ho_{0.25}Al_2$ was grown by Czochralski method in a tetra-arc furnace. The crystal is pulled at a linear speed of ~ 10 mm/hr for ~ 8 hours. The cubic Laves phase (*C15*) of the crystal was verified using Laue x-ray diffractometer. The crystal was oriented along the crystallographic axis [100] and cut by spark erosion cutting machine. The isofield and isothermal magnetization data were obtained using Quantum Design Inc. SQUID-Vibrating Sample Magnetometer (Model S-VSM). The nominal zero field was ascertained to be in the range of 0 to 10 Oe and it could be tuned to ± 1 Oe in the S-VSM system. The *M-H* loops were typically recorded by cooling the sample first from the paramagnetic state to a selected temperature in a chosen high field value, thereafter, the *M-H* loops were traced by repeated cycling of applied field between predetermined ± $H_{max}$ values.

**Results**

Figures 1(a) to 1(c) show the thermomagnetic curves in the polycrystalline alloys of $Sm_{0.98}Gd_{0.02}Al_2$ and $Pr_{0.8}Gd_{0.2}Al_2$ and a single crystal $Nd_{0.75}Ho_{0.25}Al_2$, respectively. The nominal zero field cooled (ZFC) *M-T* curve can be compared with the (high) in-field magnetization data in each of the alloys. The respective magnetic ordering temperatures ($T_c$) can be easily identified in the ZFC *M-T* data in different panels of Fig. 1. Each of the ZFC *M-H* curves crosses the *M* = 0 axis and moves towards the negative metastable values at a characteristic magnetic compensation temperature ($T_{comp}$), as identified in Figs. 1(a) to 1(c). The *M-T* curves at high fields ($H \geq 10$ kOe) in Fig. 1(a) to 1(c) show characteristic turnaround behaviour at temperatures, which are just above the corresponding $T_{comp}$ values. As stated earlier, the crossover of the *M* = 0 axis at $T_{comp}$, and a minimum in high field M-T response at $T \sim T_{comp}$ are the hallmarks of magnetic compensation phenomenon involving competing contributions to the magnetization response.

We measured the *M-H* loops at different temperatures in the magnetically ordered states of all the three samples, following the procedure described earlier. We have chosen to illustrate the important results of the magnetization hysteresis behavior in Figs. 2 to 4. Figure 2 shows the portions of the *M-H* loops over ± 30 kOe in $Sm_{0.98}Gd_{0.02}Al_2$ at *T* = 65 K, 72 K, 82 K and 88 K. The *M-H* loops display the collapse of the hysteresis response at $T_{comp} \approx 82$ K. Figures 3 and 4 display the magnetic hysteresis loops in the neighborhood of $T_{comp}$ in $Sm_{0.98}Gd_{0.02}Al_2$ and $Nd_{0.75}Ho_{0.25}Al_2$, respectively. We would like to highlight three important observations in these two figures:

(i) An inset in Fig. 3 shows a comparison of the portions of the *M-H* loops over ± 37.5 kOe at *T* = 81 K, 82 K and 84 K in $Sm_{0.98}Gd_{0.02}Al_2$; the quasi-linear behaviour (with tiny residual hysteresis) in *M-H* response in all the three curves attest to the quasi-antiferromagnetic behaviour closer to $T_{comp}$ values in this alloy (see also inset (b) in Fig. 4 for the data near $T_{comp}$ in $Nd_{0.75}Ho_{0.25}Al_2$ sample).

(ii) Figure 3 clearly shows that the width of the hysteresis loop is minimal at $T_{comp}$ of 82 K, and a few K away on the either side, the width rapidly increases.

(iii) The main panel in Fig. 4 shows the portions of the *M-H* loops over ± 800 Oe in the single crystal of $Nd_{0.75}Ho_{0.25}Al_2$ for *H* ∥ [100] at *T* = 23.75 K and *T* = 24 K. It is well apparent that the *M-H* loops are asymmetric and the mid-points of the loops at 24 K and 23.75 K are respectively located towards the left and the right of *H* = 0. The inset panel (a) in Fig. 4 shows a portion of the *M-H* loop over ± 800 Oe at an intermediate temperature, *T* = 23.85K, this loop indeed appears symmetric w.r.t. the origin.

From the *M-H* loops over the entire temperature interval from 5 K to the nominal ordering temperature of each of the alloy, we determined the values of effective coercive field (half-width of the *M-H* loop, $H_c^{eff} = (H_+ - H_-) / 2$) and the exchange bias field defined as, $H_{exch} = -(H_+ + H_-)$, where $H_+$ and $H_-$ refer to the fields at which *M-H* loop crosses the $M = 0$ axis (see Fig. 3). The panels (a) to (f) in Fig. 5 present a collation of the values of $H_{exch}$ and $H_c^{eff}$ as a function of temperature in the three alloys under study. In Fig. 5(a), the minimum in the $H_c^{eff}$ is at $T_{comp}$ = 82 K, whereas on approaching the $T_{comp}$ from 60 K and 100 K, the $H_c^{eff}$ values first appear to diverge, but, they eventually collapse towards a minimum value, while embracing the $T_{comp}$ of 82 K. The temperature variation of $H_{exch}(T)$ in Fig. 5(b) reveals the surfacing of asymmetry in the *M-H* loop as the $H_c^{eff}$ values start to decrease on approaching very close to $T_{comp}$ of 82 K in $Sm_{0.98}Gd_{0.02}Al_2$. The change in the sign of the $H_{exch}$ across $T_{comp}$ is a very noteworthy characteristic. The comparison of the $H_c^{eff}(T)$ curves in the panels (c) and (d) with the corresponding $H_{exch}(T)$ curves in the panels (e) and (f) of Fig. 5 in samples of $Pr_{0.8}Gd_{0.2}Al_2$ and $Nd_{0.75}Ho_{0.25}Al_2$, respectively bring out the similarities in the magnetic response of these samples *vis a vis* that in the much investigated $Sm_{0.98}Gd_{0.02}Al_2$ alloy. The reversal in the sign of the exchange field across $T_{comp}$ and the collapse of the effective coercive field at $T_{comp}$ are likely to be the generic features of 'ferromagnets with no net magnetic moment' [11], which imbibe magnetic compensation phenomenon in admixed ferromagnetic series of rare earth alloys.

**Discussion**

The admixed stoichiometries, $Sm_{0.98}Gd_{0.02}Al_2$, $Pr_{0.8}Gd_{0.2}Al_2$ and $Nd_{0.75}Ho_{0.25}Al_2$, which have been shown to display identical features in magnetic compensation phenomenon and reversal in sign of exchange bias field in Fig. 1 and Fig. 5 have subtle differences amongst them, as regards their $T_c$ and $T_{comp}$ values. SmAl$_2$ and PrAl$_2$ have magnetic moments per formula unit of 0.2 $\mu_B$ and 2.8 $\mu_B$ and $T_c$ values of 125 K and 40 K [18], respectively. GdAl$_2$ has T$_c$ of 170 K and magnetic moment per formula unit in it is 7.7 $\mu_B$ [18]. While SmAl$_2$ requires only 1 to 3 at. % substitution of Sm$^{3+}$ by Gd$^{3+}$ ion to elucidate magnetic compensation, in PrAl$_2$, one needs to replace 15 to 20 at. % of Pr$^{3+}$ by Gd$^{3+}$ to yield magnetic compensation behaviour. The $T_c$ of Sm$_{0.98}$Gd$_{0.02}$Al$_2$ is almost the same as that in SmAl$_2$, and the magnetic compensation in it occurs much further down at 82 K. The T$_c$ in Pr$_{0.8}$Gd$_{0.2}$Al$_2$ enhances to 64 K, as compared to that of 40 K in pure PrAl$_2$, and the magnetic compensation in it is observed at 38 K. The $T_c$ in the Nd$_{0.75}$Ho$_{0.25}$Al$_2$ is once again close to that in the pure NdAl$_2$ ($\mu_B$ / f.u. ≈ 2.5 $\mu_B$) compound, and the magnetic compensation in it gets observed at 24 K, a temperature value lower than the $T_c$ of 35 K in pure HoAl$_2$ ($\mu_B$ / f.u. ≈ 8 $\mu_B$).

Magnetic compensation behaviour in $R_{1-x}R'_x$Al$_2$ alloys is rationalized in terms of different temperature dependences of $\mu_R$ and $\mu_{R'}$, even while $R$ and $R'$ occupy the same crystallographic site in the matrix. In recent years, the magnetic compensation behaviour in the Gd doped SmAl$_2$ matrix has been described in terms of a competition between orbital and spin parts of Sm$^{3+}$ ions, which is specific to the crystalline and exchange field induced admixture effects between the ground state and the low lying first excited state of Sm$^{3+}$ [11,19,20]. The admixture effects result in inducing different temperature dependences for spin and orbital contributions to the magnetic moment of Sm$^{3+}$. The similarity in the behaviour of Sm$_{0.98}$Gd$_{0.02}$Al$_2$ and the other two alloys imply that specific attributes of Sm$^{3+}$ ions are not necessary to elucidate the magnetic compensation behaviour (cf. Fig. 1) and the associated characteristics (cf. Fig. 5).

The observed modulations in the temperature dependences of effective coercive field $H_c^{eff}$ and the notional exchange field $H_{exch}$ in Fig. 5 need an understanding in terms of the antiferromagnetic coupling between the magnetic moments of dissimilar rare earth ions belonging to the two halves of RE series ($S$-state Gd$^{3+}$ ions, for which $J = S$, are to be counted alongwith the second half RE ions, for whom $J = L + S$). It is fruitful to recall here that Webb *et al.* [16] had reported the observation of collapse in the effective coercive field [Fig.7, Ref.16] as well as divergence [Fig. 8, Ref.16] in the effective coercive field as $T \to T_{comp}$ in different co-deposited Gd-Co samples in the thin film form. First of these samples was adjudged as a uniform amorphous ferrimagnet, whereas, the other showing divergence in $H_c^{eff}$ was termed as an inhomogeneous or a macroscopic ferrimagnet, as the competing components of its magnetic response were physically separated over dimensions larger than an atomic unit cell.

In another multilayered composite comprising $GdCo_2$/Co, Webb *et al.* [21] reported the observations of the divergence and the sign change in $H_{exch}(T)$ plot as $T \to T_{comp}$ of this composite from either end [see Fig. 2, Ref. 21]. This sample was surmised to display divergence(s) in the effective coercive field as well as collapse in it at $T = T_{comp}$. In the multilayered $GdCo_2$/Co composite, the antiferromagnetically coupled Gd and Co moments in the $GdCo_2$ part are considered to imbibe the magnetic compensation characteristic, and the ferromagnetic Co layers are interfacially linked to it. It was argued by Webb *et. al.* [21] that the interfacial exchange field relates inversely to the difference in sublattice magnetizations of Gd and Co moments. In view of the fact that 'difference magnetization' signal in a quasi-ferrimagnet approaches zero at $T = T_{comp}$, and undergoes a phase reversal across it, the $H_{exch}$ in such a system would show divergence(s) in positive (negative) directions.

In the case of $R_{1-x}R'Al_2$ alloys in panels (b), (d) and (f) of Fig. 5, the notional divergence(s) and phase reversal in $H_{exch}(T)$ on going across $T_{comp}$ values do appear to find an echo in the behaviour described above for $GdCo_2$/Co compound. However, the $H_c^{eff}(T)$ data in panels (a), (c) and (e) in Fig. 5 imbibe both the characteristics, viz., the divergence as well as the collapse, as the temperature region across $T_{comp}$ is traversed. It is difficult to easily delineate the counterparts of ferromagnetic Co layer and the ferrimagnetic $GdCo_2$ stoichiometry of the $GdCo_2$/Co composite in the $R_{1-x}R'Al_2$ alloys. The polarized conduction electrons remain exchange coupled to the 4*f*-spins of both the RE ions in steadfast manner at all temperatures. Considering that the magnetic moments of *R* and *R'* are antiferromagnetically coupled, the magnetic contribution from polarized conduction electron would be in phase with the magnetization contribution from only one of the RE ions above $T_{comp}$. As the magnetic moments of the dissimilar RE ions reverse orientation across $T_{comp}$, the magnetization contribution from the polarized conduction electrons to the net magnetization signal would also reverse sign. If the conduction electrons assume the role of a soft (unidirectional anisotropy) ferromagnet in $R_{1-x}R'$Al$_2$ aloys, then the antiferromagnetically coupled local moments of the $R$ and $R'$ can be considered to assume the role of pseudo-antiferromagnet near $T_{comp}$. Such a naive picture can in principle, rationalize the surfacing of the $H_{exch}$ field as $T_{comp}$ is approached. The reversal in the sign of $H_{exch}$ across $T_{comp}$ may have some correspondence with the reversal in the orientation of conduction electron polarization w.r.t. the direction of the externally applied field. Between $T_c$ and $T_{comp}$, the (soft) conduction electron contribution is aligned in the direction of external field, whereas below $T_{comp}$, this contribution reverses its sign giving rise to an anomalous situation of the soft part being antiparallel to the external field.

The minimum in the width of the hysteresis loop at $T = T_{comp}$ may be accounted by asserting that in an ideal antiferromagnet, magnetization response is expected to approach zero as the applied field is removed. If the divergence in the coercive field is an attribute anticipated [16] in inhomogeneous ferrimagnets, one could argue that the admixed rare earth alloys displaying magnetic compensation behaviour also belong to the same category. Multi-layer composites involving all types of ferrimagnets, homogeneous or inhomogeneous, are argued [16] to yield collapse in the effective coercive field, whose proximity to $T_{comp}$ (within 1% of $(T - T_{comp}) / T_{comp}$) depends on the strength of coupling across ferro-antiferro interface. We believe that the temperature intervals of the collapse in $H_c^{eff}$ displayed in the panels of Fig. 5 are in the right ball park, and such a behavior is an exemplification of the divergence and the collapse, articulated in Ref. [16].

**Summary and Conclusion**

The light (first-half) - heavy (second-half) combination of 4*f*-rare earths and the non-magnetic elements forming admixed intermetallic alloys display the notion of '4*f*-spin ferromagnetism' and the antiparallel alignment of the magnetic moments of dissimilar rare earth elements governed by the indirect RKKY interaction between the 4f-spins of the rare earth ions. A suitable choice of two rare earth elements, based on their magnetic moments, can drive a given alloy towards a state of near-zero magnetization in its thermomagnetic response. Above a threshold magnetic field, the magnetic moments of rare earth elements undergo a reversal process, seen as a turnaround in the thermomagnetic response across the magnetic compensation temperature. We have reported the observation of the presence of the exchange bias field in the vicinity of the magnetic compensation temperature, with opposite signs below and above the compensation temperature in three specific samples of admixed $R_{1-x}R'$Al$_2$ alloys, viz., Sm$_{0.98}$Gd$_{0.02}$Al$_2$ (polycrystalline), Pr$_{0.8}$Gd$_{0.2}$Al$_2$ (polycrystalline) and Nd$_{0.75}$Ho$_{0.25}$Al$_2$ (single crystal).

The presence of a large exchange bias field in the reported alloys, alongwith their near-zero magnetization response of these stoichiometries could have potential for applications in the area of Spintronics, where the thin film form of these alloys sandwitched between the magnetic multilayers could help in tuning the magnetic properties of the devices. One of the applications could be to fabricate a multilayer structure comprising 'ferromagnet with no net magnetization' (as an AF layer) and the transition metal (or rare earth based) alloy as a ferromagnet. Below the nominal ordering temperature of the said ferromagnets having no net magnetization (Neel temperature in the traditional sense), the interface exchange bias will be experienced by the transition metal ferromagnetic layer. However, when the 'no net magnetization' alloy would get cooled close to the magnetic compensation temperature, an additional exchange bias could be generated which will either add or subtract to the already present bias field, just above or below the magnetic compensation temperature, respectively. This could provide the possibility of tailoring the magnetoresistance of the multilayered structures.

**Acknowledgements**

We thank Mr. Raghav Mohta for his participation in some of the explorations in Sm(Gd)Al$_2$ and Mr. Ravi Singh for discussions on exchange bias phenomenon.

**FIGURE CAPTIONS**

Fig. 1. Temperature variation of the field cooled magnetization in (a) $Sm_{0.98}Gd_{0.02}Al_2$, (b) $Pr_{0.8}Gd_{0.2}Al_2$ and (c) $Nd_{0.75}Ho_{0.25}Al_2$ in nominal zero field (ZFC curves) and in high fields (10 to 14 kOe) at temperatures marked as compensation temperatures ($T_{comp}$). At higher values of applied magnetic field, a turnaround in the thermomagnetic curve is seen due to reversal in the orientations of the magnetic moments of the dissimilar rare earth ions. $T_c$ value is also marked in the ZFC curve of each of the alloy.

Fig. 2. Magnetic hysteresis loops in $Sm_{0.98}Gd_{0.02}Al_2$ at four different temperatures, as indicated in the figure. The *M-H* loops at 65 K and 72 K are below $T_{comp}$ and that at 88 K is above $T_{comp}$. The width of the loop dramatically shrinks at the magnetic compensation temperature, $T_{comp} = 82$ K.

Fig. 3. Magnetic hysteresis loops in $Sm_{0.98}Gd_{0.02}Al_2$ at 81 K, 82 K and 84 K. The width of the *M-H* loop is minimum at $T_{comp} = 82$ K and the loop is symmetric. The *M-H* loops at 81 K and 84 K, however, have an asymmetric nature. The inset panel in Fig. 3 shows the *M-H* loops over a range of ± 40 kOe.

Fig. 4. Portions of the *M-H* loops over ± 800 Oe in $Nd_{0.75}Ho_{0.25}Al_2$ at 23.75 K and 24 K. The left and the right shift of the *M-H* loops above ($T = 24$ K) and below ($T = 23.75$ K) the magnetic compensation temperature, respectively can be clearly seen in the main panel. The inset panel (b) in Fig.4 shows the *M-H* loops in the field range of ± 10 kOe at T = 23.75 K and 24 K. The linear part in the *M* versus *H* loops at high fields indicates an antiparallel alignment of the rare earth moments. The inset panel (a) in Fig. 4 shows the portion of the *M-H* loop over ± 800 Oe at $T_{comp} = 23.85$ K.

Fig. 5. Temperature variation of the effective coercive field and the exchange bias field in samples of $Sm_{0.98}Gd_{0.02}Al_2$ (panels a & b), $Pr_{0.8}Gd_{0.2}Al_2$ (panels c & d) and $Nd_{0.75}Ho_{0.25}Al_2$ (panels e & f). For all the three alloys, the effective coercive field collapses at the respective magnetic compensation temperature and the exchange bias fields change sign, while going across it.

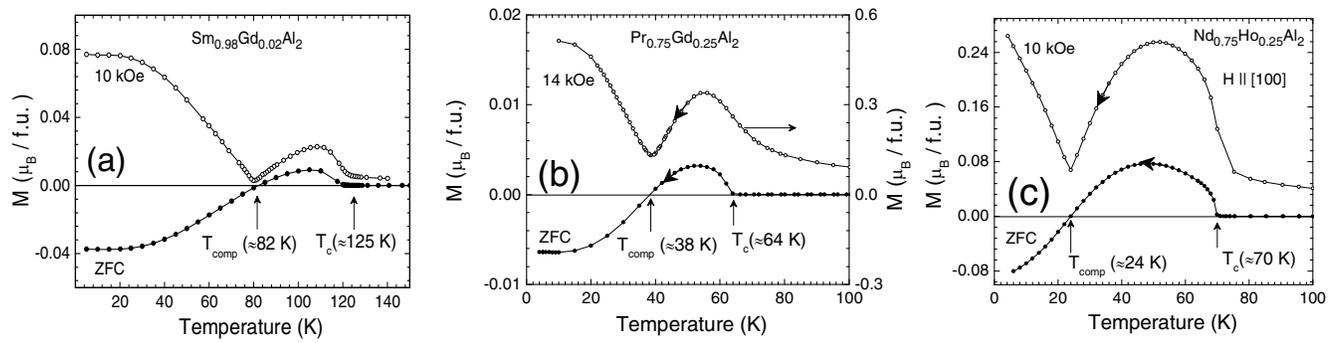

Fig. 1.

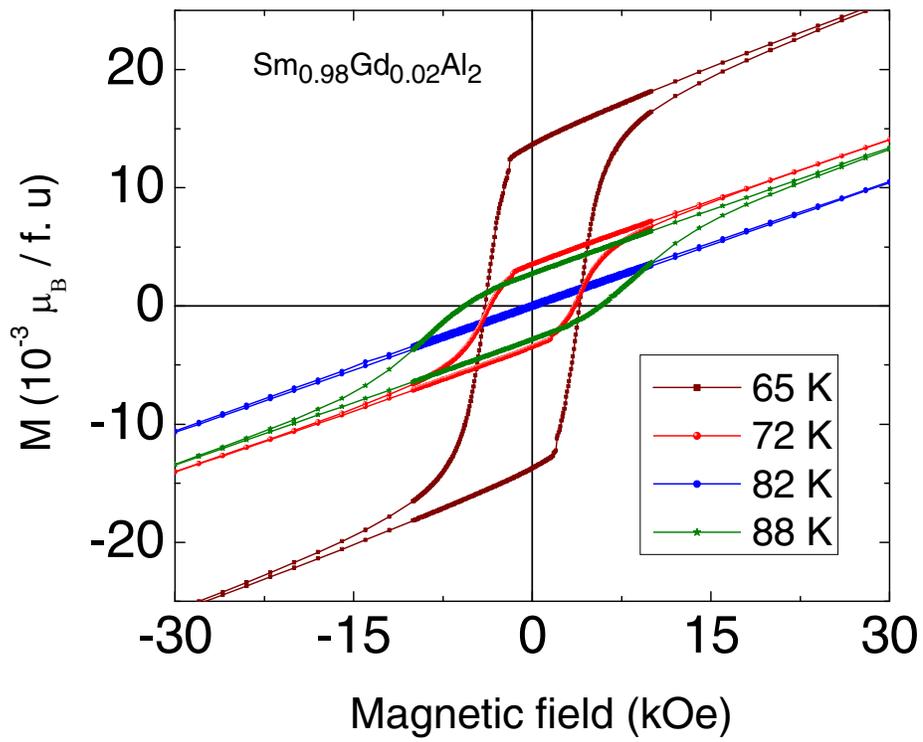

Fig. 2.

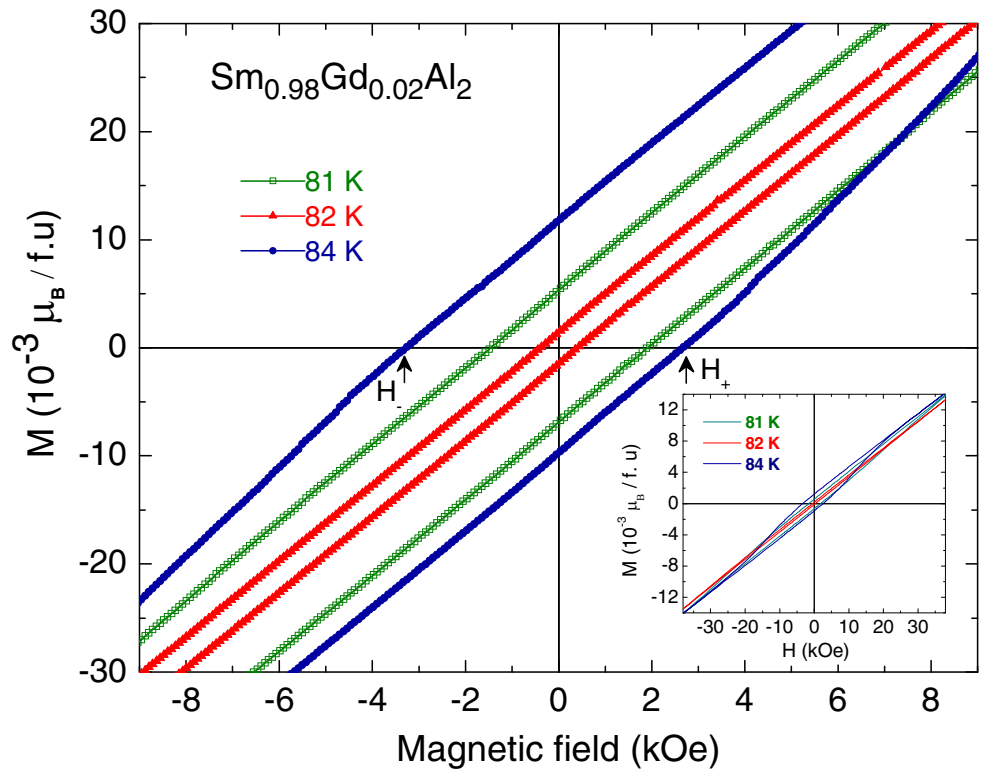

Fig. 3.

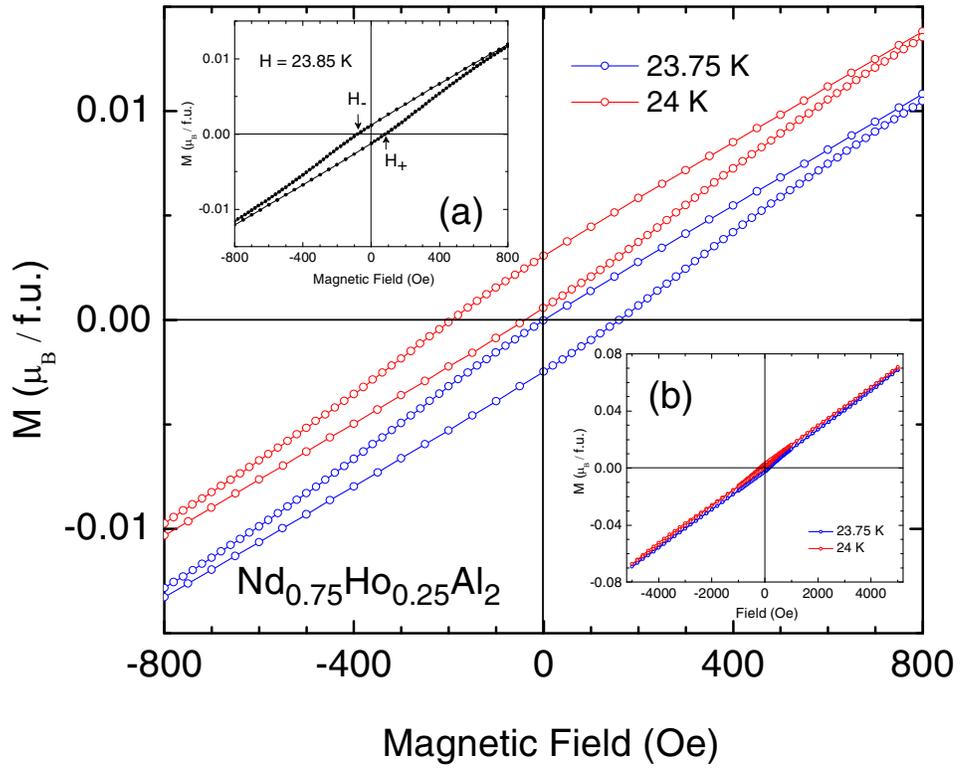

Fig. 4.

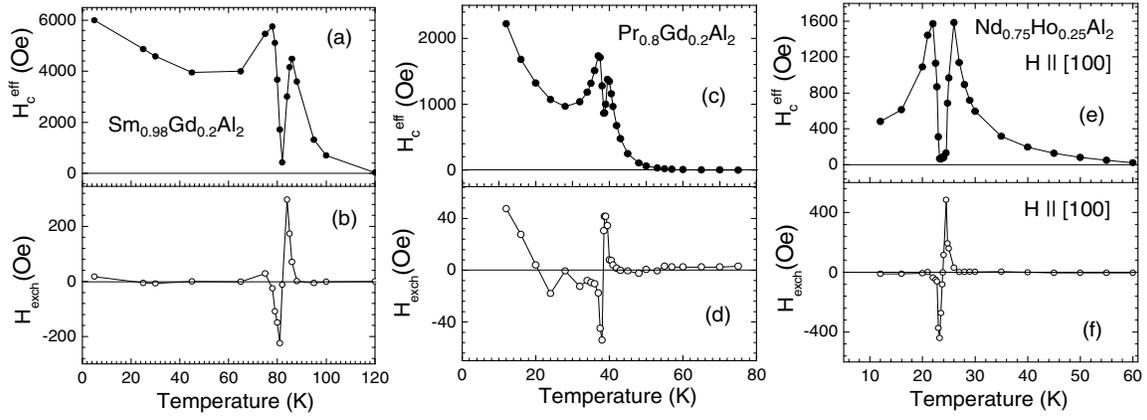

Fig. 5.